\documentclass[10pt, letterpaper, conference]{IEEEtran} 
\IEEEoverridecommandlockouts
\usepackage{cite}
\usepackage{comment}
\usepackage{amsmath,amssymb,amsfonts}
\usepackage{algorithmic}
\usepackage{graphicx}
\usepackage{float}
\usepackage{caption}
\usepackage{subcaption}
 \usepackage{svg}
\usepackage{subfig}
\usepackage{textcomp}
\usepackage{multirow}
\usepackage{xcolor}
\def\BibTeX{{\rm B\kern-.05em{\sc i\kern-.025em b}\kern-.08em
    T\kern-.1667em\lower.7ex\hbox{E}\kern-.125emX}}

\usepackage[acronyms,nonumberlist,nopostdot,nomain,nogroupskip,acronymlists={hidden}]{glossaries}
\newacronym{3gpp}{3GPP}{3rd Generation Partnership Project}
\newacronym{4g}{4G}{4th generation}
\newacronym{5g}{5G}{5th generation}
\newacronym{6g}{6G}{6th generation}
\newacronym{5gc}{5GC}{5G Core}
\newacronym{aau}{AAU}{Active Antenna Unit}
\newacronym{adc}{ADC}{Analog to Digital Converter}
\newacronym{aerpaw}{AERPAW}{Aerial Experimentation and Research Platform for Advanced Wireless}
\newacronym{ai}{AI}{Artificial Intelligence}
\newacronym{aimd}{AIMD}{Additive Increase Multiplicative Decrease}
\newacronym{am}{AM}{Acknowledged Mode}
\newacronym{amc}{AMC}{Adaptive Modulation and Coding}
\newacronym{amf}{AMF}{Access and Mobility Management Function}
\newacronym{aops}{AOPS}{Adaptive Order Prediction Scheduling}
\newacronym{api}{API}{Application Programming Interface}
\newacronym{apn}{APN}{Access Point Name}
\newacronym{ap}{AP}{Application Protocol}
\newacronym{aqm}{AQM}{Active Queue Management}
\newacronym{ausf}{AUSF}{Authentication Server Function}
\newacronym{avc}{AVC}{Advanced Video Coding}
\newacronym{awgn}{AGWN}{Additive White Gaussian Noise}
\newacronym{balia}{BALIA}{Balanced Link Adaptation Algorithm}
\newacronym{bbu}{BBU}{Base Band Unit}
\newacronym{bdp}{BDP}{Bandwidth-Delay Product}
\newacronym{ber}{BER}{Bit Error Rate}
\newacronym{bf}{BF}{Beamforming}
\newacronym{bler}{BLER}{Block Error Rate}
\newacronym{brr}{BRR}{Bayesian Ridge Regressor}
\newacronym{bs}{BS}{Base Station}
\newacronym{bsr}{BSR}{Buffer Status Report}
\newacronym{bss}{BSS}{Business Support System}
\newacronym{ca}{CA}{Carrier Aggregation}
\newacronym{caas}{CaaS}{Connectivity-as-a-Service}
\newacronym{cav}{CAV}{Connected and Autonoums Vehicle}
\newacronym{cb}{CB}{Code Block}
\newacronym{cc}{CC}{Congestion Control}
\newacronym{ccid}{CCID}{Congestion Control ID}
\newacronym{cco}{CC}{Carrier Component}
\newacronym{cd}{CD}{Continuous Delivery}
\newacronym{cdd}{CDD}{Cyclic Delay Diversity}
\newacronym{cdf}{CDF}{Cumulative Distribution Function}
\newacronym{cdn}{CDN}{Content Distribution Network}
\newacronym{cli}{CLI}{Command-line Interface}
\newacronym{cn}{CN}{Core Network}
\newacronym{codel}{CoDel}{Controlled Delay Management}
\newacronym{comac}{COMAC}{Converged Multi-Access and Core}
\newacronym{cord}{CORD}{Central Office Re-architected as a Datacenter}
\newacronym{cornet}{CORNET}{COgnitive Radio NETwork}
\newacronym{cosmos}{COSMOS}{Cloud Enhanced Open Software Defined Mobile Wireless Testbed for City-Scale Deployment}
\newacronym{cots}{COTS}{Commercial Off-the-Shelf}
\newacronym{cp}{CP}{Control Plane}
\newacronym{cyp}{CP}{Cyclic Prefix}
\newacronym{up}{UP}{User Plane}
\newacronym{cpu}{CPU}{Central Processing Unit}
\newacronym{cqi}{CQI}{Channel Quality Information}
\newacronym{cr}{CR}{Cognitive Radio}
\newacronym{cran}{CRAN}{Cloud \gls{ran}}
\newacronym{crs}{CRS}{Cell Reference Signal}
\newacronym{csi}{CSI}{Channel State Information}
\newacronym{csirs}{CSI-RS}{Channel State Information - Reference Signal}
\newacronym{cu}{CU}{Central Unit}
\newacronym{dac}{DAC}{Digital to Analog Converter}
\newacronym{dag}{DAG}{Directed Acyclic Graph}
\newacronym{das}{DAS}{Distributed Antenna System}
\newacronym{dash}{DASH}{Dynamic Adaptive Streaming over HTTP}
\newacronym{dc}{DC}{Dual Connectivity}
\newacronym{dccp}{DCCP}{Datagram Congestion Control Protocol}
\newacronym{dce}{DCE}{Direct Code Execution}
\newacronym{dci}{DCI}{Downlink Control Information}
\newacronym{dctcp}{DCTCP}{Data Center TCP}
\newacronym{dl}{DL}{Downlink}
\newacronym{dmr}{DMR}{Deadline Miss Ratio}
\newacronym{dmrs}{DMRS}{DeModulation Reference Signal}
\newacronym{drlcc}{DRL-CC}{Deep Reinforcement Learning Congestion Control}
\newacronym{dsrc}{DSRC}
{dedicated short-range communications}
\newacronym{d2d}{D2D}{device-to-device}
\newacronym{drs}{DRS}{Discovery Reference Signal}
\newacronym{du}{DU}{Distributed Unit}
\newacronym{e2e}{E2E}{end-to-end}
\newacronym{earfcn}{EARFCN}{E-UTRA Absolute Radio Frequency Channel Number}
\newacronym{ecaas}{ECaaS}{Edge-Cloud-as-a-Service}
\newacronym{ecn}{ECN}{Explicit Congestion Notification}
\newacronym{edf}{EDF}{Earliest Deadline First}
\newacronym{embb}{eMBB}{Enhanced Mobile Broadband}
\newacronym{empower}{EMPOWER}{EMpowering transatlantic PlatfOrms for advanced WirEless Research}
\newacronym{enb}{eNB}{evolved Node Base}
\newacronym{endc}{EN-DC}{E-UTRAN-\gls{nr} \gls{dc}}
\newacronym{epc}{EPC}{Evolved Packet Core}
\newacronym{eps}{EPS}{Evolved Packet System}
\newacronym{es}{ES}{Edge Server}
\newacronym{etsi}{ETSI}{European Telecommunications Standards Institute}
\newacronym[firstplural=Estimated Times of Arrival (ETAs)]{eta}{ETA}{Estimated Time of Arrival}
\newacronym{eutran}{E-UTRAN}{Evolved Universal Terrestrial Access Network}
\newacronym{faas}{FaaS}{Function-as-a-Service}
\newacronym{fapi}{FAPI}{Functional Application Platform Interface}
\newacronym{fdd}{FDD}{Frequency Division Duplexing}
\newacronym{fdm}{FDM}{Frequency Division Multiplexing}
\newacronym{fdma}{FDMA}{Frequency Division Multiple Access}
\newacronym{fed4fire}{FED4FIRE+}{Federation 4 Future Internet Research and Experimentation Plus}
\newacronym{fir}{FIR}{Finite Impulse Response}
\newacronym{fit}{FIT}{Future \acrlong{iot}}
\newacronym{fpga}{FPGA}{Field Programmable Gate Array}
\newacronym{fr2}{FR2}{Frequency Range 2}
\newacronym{fr1}{FR1}{Frequency Range 1}
\newacronym{fs}{FS}{Fast Switching}
\newacronym{fscc}{FSCC}{Flow Sharing Congestion Control}
\newacronym{ftp}{FTP}{File Transfer Protocol}
\newacronym{fw}{FW}{Flow Window}
\newacronym{ge}{GE}{Gaussian Elimination}
\newacronym{gnb}{gNB}{Next Generation NodeB}
\newacronym{gop}{GOP}{Group of Pictures}
\newacronym{gpr}{GPR}{Gaussian Process Regressor}
\newacronym{gpu}{GPU}{Graphics Processing Unit}
\newacronym{gtp}{GTP}{GPRS Tunneling Protocol}
\newacronym{gtpc}{GTP-C}{GPRS Tunnelling Protocol Control Plane}
\newacronym{gtpu}{GTP-U}{GPRS Tunnelling Protocol User Plane}
\newacronym{gw}{GW}{Gateway}
\newacronym{harq}{HARQ}{Hybrid Automatic Repeat reQuest}
\newacronym{hetnet}{HetNet}{Heterogeneous Network}
\newacronym{hh}{HH}{Hard Handover}
\newacronym{hol}{HOL}{Head-of-Line}
\newacronym{hqf}{HQF}{Highest-quality-first}
\newacronym{hss}{HSS}{Home Subscription Server}
\newacronym{http}{HTTP}{HyperText Transfer Protocol}
\newacronym{ia}{IA}{Initial Access}
\newacronym{iab}{IAB}{Integrated Access and Backhaul}
\newacronym{ic}{IC}{Incident Command}
\newacronym{ietf}{IETF}{Internet Engineering Task Force}
\newacronym{imsi}{IMSI}{International Mobile Subscriber Identity}
\newacronym{imt}{IMT}{International Mobile Telecommunication}
\newacronym{iot}{IoT}{Internet of Things}
\newacronym{ip}{IP}{Internet Protocol}
\newacronym{itu}{ITU}{International Telecommunication Union}
\newacronym{kpi}{KPI}{Key Performance Indicator}
\newacronym{kpm}{KPM}{Key Performance Measurement}
\newacronym{kvm}{KVM}{Kernel-based Virtual Machine}
\newacronym{los}{LoS}{Line of Sight}
\newacronym{lsm}{LSM}{Link-to-System Mapping}
\newacronym{lstm}{LSTM}{Long Short Term Memory}
\newacronym{lte}{LTE}{Long Term Evolution}
\newacronym{lxc}{LXC}{Linux Container}
\newacronym{m2m}{M2M}{Machine to Machine}
\newacronym{mac}{MAC}{Medium Access Control}
\newacronym{manet}{MANET}{Mobile Ad Hoc Network}
\newacronym{mano}{MANO}{Management and Orchestration}
\newacronym{mc}{MC}{Multi-Connectivity}
\newacronym{mcc}{MCC}{Mobile Cloud Computing}
\newacronym{mchem}{MCHEM}{Massive Channel Emulator}
\newacronym{mcs}{MCS}{Modulation and Coding Scheme}
\newacronym{mec2}{MEC}{Multi-access Edge Computing}
\newacronym{mec}{MEC}{Mobile Edge Computing}
\newacronym{mfc}{MFC}{Mobile Fog Computing}
\newacronym{mgen}{MGEN}{Multi-Generator}
\newacronym{mi}{MI}{Mutual Information}
\newacronym{mib}{MIB}{Master Information Block}
\newacronym{miesm}{MIESM}{Mutual Information Based Effective SINR}
\newacronym{mimo}{MIMO}{Multiple Input, Multiple Output}
\newacronym{ml}{ML}{Machine Learning}
\newacronym{mlr}{MLR}{Maximum-local-rate}
\newacronym[plural=\gls{mme}s,firstplural=Mobility Management Entities (MMEs)]{mme}{MME}{Mobility Management Entity}
\newacronym{mmtc}{mMTC}{Massive Machine-Type Communications}
\newacronym{mmwave}{mmWave}{millimeter wave}
\newacronym{mpdccp}{MP-DCCP}{Multipath Datagram Congestion Control Protocol}
\newacronym{mptcp}{MPTCP}{Multipath TCP}
\newacronym{mr}{MR}{Maximum Rate}
\newacronym{mrdc}{MR-DC}{Multi \gls{rat} \gls{dc}}
\newacronym{mse}{MSE}{Mean Square Error}
\newacronym{mss}{MSS}{Maximum Segment Size}
\newacronym{mt}{MT}{Mobile Termination}
\newacronym{mtd}{MTD}{Machine-Type Device}
\newacronym{mtu}{MTU}{Maximum Transmission Unit}
\newacronym{mumimo}{MU-MIMO}{Multi-user \gls{mimo}}
\newacronym{mvno}{MVNO}{Mobile Virtual Network Operator}
\newacronym{nalu}{NALU}{Network Abstraction Layer Unit}
\newacronym{nas}{NAS}{Network Attached Storage}
\newacronym{nat}{NAT}{Network Address Translation}
\newacronym{nbiot}{NB-IoT}{Narrow Band IoT}
\newacronym{nfv}{NFV}{Network Function Virtualization}
\newacronym{nfvi}{NFVI}{Network Function Virtualization Infrastructure}
\newacronym{ni}{NI}{Network Interfaces}
\newacronym{nic}{NIC}{Network Interface Card}
\newacronym{now}{NOW}{Non Overlapping Window}
\newacronym{nsm}{NSM}{Network Service Mesh}
\newacronym{nr}{NR}{New Radio}
\newacronym{nrf}{NRF}{Network Repository Function}
\newacronym{nsa}{NSA}{Non Stand Alone}
\newacronym{nse}{NSE}{Network Slicing Engine}
\newacronym{nssf}{NSSF}{Network Slice Selection Function}
\newacronym{o2i}{O2I}{Outdoor to Indoor}
\newacronym{oai}{OAI}{OpenAirInterface}
\newacronym{oaicn}{OAI-CN}{\gls{oai} \acrlong{cn}}
\newacronym{oairan}{OAI-RAN}{\acrlong{oai} \acrlong{ran}}
\newacronym{oam}{OAM}{Operations, Administration and Maintenance}
\newacronym{ofdm}{OFDM}{Orthogonal Frequency Division Multiplexing}
\newacronym{olia}{OLIA}{Opportunistic Linked Increase Algorithm}
\newacronym{omec}{OMEC}{Open Mobile Evolved Core}
\newacronym{onap}{ONAP}{Open Network Automation Platform}
\newacronym{onf}{ONF}{Open Networking Foundation}
\newacronym{onos}{ONOS}{Open Networking Operating System}
\newacronym{oom}{OOM}{\gls{onap} Operations Manager}
\newacronym{opnfv}{OPNFV}{Open Platform for \gls{nfv}}
\newacronym{oran}{O-RAN}{Open \gls{ran}}
\newacronym{orbit}{ORBIT}{Open-Access Research Testbed for Next-Generation Wireless Networks}
\newacronym{os}{OS}{Operating System}
\newacronym{oss}{OSS}{Operations Support System}
\newacronym{pa}{PA}{Position-aware}
\newacronym{pase}{PASE}{Prioritization, Arbitration, and Self-adjusting Endpoints}
\newacronym{pawr}{PAWR}{Platforms for Advanced Wireless Research}
\newacronym{pbch}{PBCH}{Physical Broadcast Channel}
\newacronym{pcef}{PCEF}{Policy and Charging Enforcement Function}
\newacronym{pcfich}{PCFICH}{Physical Control Format Indicator Channel}
\newacronym{pcrf}{PCRF}{Policy and Charging Rules Function}
\newacronym{pdcch}{PDCCH}{Physical Downlink Control Channel}
\newacronym{pdcp}{PDCP}{Packet Data Convergence Protocol}
\newacronym{pdsch}{PDSCH}{Physical Downlink Shared Channel}
\newacronym{pdu}{PDU}{Packet Data Unit}
\newacronym{pf}{PF}{Proportional Fair}
\newacronym{pgw}{PGW}{Packet Gateway}
\newacronym{phich}{PHICH}{Physical Hybrid ARQ Indicator Channel}
\newacronym{phy}{PHY}{Physical}
\newacronym{pmch}{PMCH}{Physical Multicast Channel}
\newacronym{pmi}{PMI}{Precoding Matrix Indicators}
\newacronym{powder}{POWDER}{Platform for Open Wireless Data-driven Experimental Research}
\newacronym{ppo}{PPO}{Proximal Policy Optimization}
\newacronym{ppp}{PPP}{Poisson Point Process}
\newacronym{prach}{PRACH}{Physical Random Access Channel}
\newacronym{prb}{PRB}{Physical Resource Block}
\newacronym{psnr}{PSNR}{Peak Signal to Noise Ratio}
\newacronym{pss}{PSS}{Primary Synchronization Signal}
\newacronym{pucch}{PUCCH}{Physical Uplink Control Channel}
\newacronym{pusch}{PUSCH}{Physical Uplink Shared Channel}
\newacronym{qam}{QAM}{Quadrature Amplitude Modulation}
\newacronym{qci}{QCI}{\gls{qos} Class Identifier}
\newacronym{qoe}{QoE}{Quality of Experience}
\newacronym{qos}{QoS}{Quality of Service}
\newacronym{quic}{QUIC}{Quick UDP Internet Connections}
\newacronym{ra}{RA}{Resouces Allocation}
\newacronym{rach}{RACH}{Random Access Channel}
\newacronym{ran}{RAN}{Radio Access Network}
\newacronym[firstplural=Radio Access Technologies (RATs)]{rat}{RAT}{Radio Access Technology}
\newacronym{rbg}{RBG}{Resource Block Group}
\newacronym{rcn}{RCN}{Research Coordination Network}
\newacronym{rc}{RC}{RAN Control}
\newacronym{rec}{REC}{Radio Edge Cloud}
\newacronym{red}{RED}{Random Early Detection}
\newacronym{renew}{RENEW}{Reconfigurable Eco-system for Next-generation End-to-end Wireless}
\newacronym{rf}{RF}{Radio Frequency}
\newacronym{rfc}{RFC}{Request for Comments}
\newacronym{rfr}{RFR}{Random Forest Regressor}
\newacronym{ric}{RIC}{\gls{ran} Intelligent Controller}
\newacronym{rlc}{RLC}{Radio Link Control}
\newacronym{rlf}{RLF}{Radio Link Failure}
\newacronym{rlnc}{RLNC}{Random Linear Network Coding}
\newacronym{rmr}{RMR}{RIC Message Router}
\newacronym{rmse}{RMSE}{Root Mean Squared Error}
\newacronym{rnis}{RNIS}{Radio Network Information Service}
\newacronym{rr}{RR}{Round Robin}
\newacronym{rrc}{RRC}{Radio Resource Control}
\newacronym{rrm}{RRM}{Radio Resource Management}
\newacronym{rru}{RRU}{Remote Radio Unit}
\newacronym{rs}{RS}{Remote Server}
\newacronym{rsrp}{RSRP}{Reference Signal Received Power}
\newacronym{rsrq}{RSRQ}{Reference Signal Received Quality}
\newacronym{rss}{RSS}{Received Signal Strength}
\newacronym{rssi}{RSSI}{Received Signal Strength Indicator}
\newacronym{rtt}{RTT}{Round Trip Time}
\newacronym{ru}{RU}{Radio Unit}
\newacronym{rus}{RSU}{Road Side Unit}
\newacronym{rw}{RW}{Receive Window}
\newacronym{rx}{RX}{Receiver}
\newacronym{s1ap}{S1AP}{S1 Application Protocol}
\newacronym{sa}{SA}{standalone}
\newacronym{sack}{SACK}{Selective Acknowledgment}
\newacronym{sap}{SAP}{Service Access Point}
\newacronym{sc2}{SC2}{Spectrum Collaboration Challenge}
\newacronym{scef}{SCEF}{Service Capability Exposure Function}
\newacronym{sch}{SCH}{Secondary Cell Handover}
\newacronym{scoot}{SCOOT}{Split Cycle Offset Optimization Technique}
\newacronym{sctp}{SCTP}{Stream Control Transmission Protocol}
\newacronym{sdap}{SDAP}{Service Data Adaptation Protocol}
\newacronym{sdk}{SDK}{Software Development Kit}
\newacronym{sdm}{SDM}{Space Division Multiplexing}
\newacronym{sdma}{SDMA}{Spatial Division Multiple Access}
\newacronym{sdn}{SDN}{Software-defined Networking}
\newacronym{sdr}{SDR}{Software-defined Radio}
\newacronym{seba}{SEBA}{SDN-Enabled Broadband Access}
\newacronym{sgsn}{SGSN}{Serving GPRS Support Node}
\newacronym{sgw}{SGW}{Service Gateway}
\newacronym{si}{SI}{Study Item}
\newacronym{sib}{SIB}{Secondary Information Block}
\newacronym{sinr}{SINR}{Signal to Interference plus Noise Ratio}
\newacronym{sip}{SIP}{Session Initiation Protocol}
\newacronym{siso}{SISO}{Single Input, Single Output}
\newacronym{sla}{SLA}{Service Level Agreement}
\newacronym{sm}{SM}{Service Model}
\newacronym{smo}{SMO}{Service Management and Orchestration}
\newacronym{smsgmsc}{SMS-GMSC}{\gls{sms}-Gateway}
\newacronym{snr}{SNR}{Signal-to-Noise-Ratio}
\newacronym{son}{SON}{Self-Organizing Network}
\newacronym{sptcp}{SPTCP}{Single Path TCP}
\newacronym{srb}{SRB}{Service Radio Bearer}
\newacronym{srn}{SRN}{Standard Radio Node}
\newacronym{srs}{SRS}{Sounding Reference Signal}
\newacronym{ss}{SS}{Synchronization Signal}
\newacronym{ssb}{SSB}{Synchronization Signal Block}
\newacronym{sss}{SSS}{Secondary Synchronization Signal}
\newacronym{st}{ST}{Spanning Tree}
\newacronym{svc}{SVC}{Scalable Video Coding}
\newacronym{tb}{TB}{Transport Block}
\newacronym{tcp}{TCP}{Transmission Control Protocol}
\newacronym{tdd}{TDD}{Time Division Duplexing}
\newacronym{tdm}{TDM}{Time Division Multiplexing}
\newacronym{tdma}{TDMA}{Time Division Multiple Access}
\newacronym{tfl}{TfL}{Transport for London}
\newacronym{tfrc}{TFRC}{TCP-Friendly Rate Control}
\newacronym{tft}{TFT}{Traffic Flow Template}
\newacronym{tgen}{TGEN}{Traffic Generator}
\newacronym{tip}{TIP}{Telecom Infra Project}
\newacronym{tm}{TM}{Transparent Mode}
\newacronym{to}{TO}{Telco Operator}
\newacronym{tr}{TR}{Technical Report}
\newacronym{trp}{TRP}{Transmitter Receiver Pair}
\newacronym{ts}{TS}{Technical Specification}
\newacronym{tti}{TTI}{Transmission Time Interval}
\newacronym{ttt}{TTT}{Time-to-Trigger}
\newacronym{tx}{TX}{Transmitter}
\newacronym{uas}{UAS}{Unmanned Aerial System}
\newacronym{uav}{UAV}{Unmanned Aerial Vehicle}
\newacronym{udm}{UDM}{Unified Data Management}
\newacronym{udp}{UDP}{User Datagram Protocol}
\newacronym{udr}{UDR}{Unified Data Repository}
\newacronym{ue}{UE}{User Equipment}
\newacronym{uhd}{UHD}{\gls{usrp} Hardware Driver}
\newacronym{ul}{UL}{Uplink}
\newacronym{um}{UM}{Unacknowledged Mode}
\newacronym{uml}{UML}{Unified Modeling Language}
\newacronym{upa}{UPA}{Uniform Planar Array}
\newacronym{upf}{UPF}{User Plane Function}
\newacronym{urllc}{URLLC}{Ultra Reliable and Low Latency Communications}
\newacronym{usa}{U.S.}{United States}
\newacronym{usim}{USIM}{Universal Subscriber Identity Module}
\newacronym{usrp}{USRP}{Universal Software Radio Peripheral}
\newacronym{utc}{UTC}{Urban Traffic Control}
\newacronym{vim}{VIM}{Virtualization Infrastructure Manager}
\newacronym{vm}{VM}{Virtual Machine}
\newacronym{vnf}{VNF}{Virtual Network Function}
\newacronym{volte}{VoLTE}{Voice over \gls{lte}}
\newacronym{voltha}{VOLTHA}{Virtual OLT HArdware Abstraction}
\newacronym{vr}{VR}{Virtual Reality}
\newacronym{vran}{vRAN}{Virtualized \gls{ran}}
\newacronym{vss}{VSS}{Video Streaming Server}
\newacronym{v2x}{V2X}{vehicle-to-everything}
\newacronym{v2i}{V2I}{vehicle-to-infrastructure}
\newacronym{v2v}{V2V}{vehicle-to-vehicle}
\newacronym{v2n}{V2N}{vehicle-to-network}
\newacronym{wbf}{WBF}{Wired Bias Function}
\newacronym{wf}{WF}{Waterfilling}
\newacronym{wg}{WG}{Working Group}
\newacronym{wlan}{WLAN}{Wireless Local Area Network}
\newacronym{osm}{OSM}{Open Source \gls{nfv} Management and Orchestration}
\newacronym{pnf}{PNF}{Physical Network Function}
\newacronym{drl}{DRL}{Deep Reinforcement Learning}
\newacronym{mtc}{MTC}{Machine-type Communications}
\newacronym{osc}{OSC}{O-RAN Software Community}
\newacronym{mns}{MnS}{Management Services}
\newacronym{ves}{VES}{\gls{vnf} Event Stream}
\newacronym{ei}{EI}{Enrichment Information}
\newacronym{fh}{FH}{Fronthaul}
\newacronym{fft}{FFT}{Fast Fourier Transform}
\newacronym{laa}{LAA}{Licensed-Assisted Access}
\newacronym{plfs}{PLFS}{Physical Layer Frequency Signals}
\newacronym{ptp}{PTP}{Precision Time Protocol}
\newacronym{lidar}{LiDAR}{Light Detection And Ranging}
\newacronym{dem}{DEM}{Digital Elevation Model}
\newacronym{dtm}{DEM}{Digital Terrain Model}
\newacronym{dsm}{DEM}{Digital Surface Models}
\newacronym{ota}{OTA}{Over-The-Air}
\newacronym{ns}{NS}{Network Slicing}
\newacronym{ne}{NE}{Nash Equilibrium}
\newacronym{hf}{HF}{High Frequency}
\newacronym{noma}{NOMA}{Non-Orthogonal Multiple Access}
\newacronym{sre}{SRE}{Smart Radio Environment}
\newacronym{ris}{RIS}{Reconfigurable Intelligent Surface}
\newacronym{inp}{InP}{Infrastructure Provider}
\newacronym{smf}{SMF}{Slicing Magangement Framework}
\newacronym{nsn}{NSN}{Network Slicing Negotiation}
\newacronym{sms}{SMS}{Slicing MAC Scheduler}
\newacronym{brd}{BRD}{Best Response Dynamics}
\newacronym{dssbr}{DSSBR}{Double Step Smoothed Best Response}
\newacronym{poa}{PoA}{Price of Anarchy}
\newacronym{pos}{PoS}{Price of Stability}
\newacronym{milp}{MILP}{Mixed Integer-Linear Program}
\newacronym{pod}{PoD}{Price of DSSBR}
\newacronym{roc}{ROC}{Radio Overload Control}
\newacronym{ciot}{cIoT}{critical Internet of Things}
\newacronym{embbpr}{eMBB Pr.}{enhanced Mobile BroadBand Premium}
\newacronym{sps}{SPS}{Semi-persistent Scheduling}
\newacronym{cg}{CG}{Configured Grant}
\newacronym{embbbs}{eMBB Bs.}{enhanced Mobile BroadBand Basic}
\newacronym{en}{EN}{Edge Node}
\newacronym{ec}{EC}{Edge Computing}
\newacronym{sp}{SP}{Service Provider}
\newacronym{me}{ME}{Market Equilibrium}
\newacronym{so}{SO}{Social Optimum}
\newacronym{wso}{WSO}{Weighted Social Optimum}
\newacronym{ps}{PS}{Proportional Sharing}
\newacronym{eg}{EG}{Eisenberg-Gale program}
\newacronym{pe}{PE}{Pareto Efficiency}
\newacronym{nsw}{NSW}{Nash Social Welfare}
\newacronym{ef}{EF}{Envy-Freeness}
\newacronym{sub6}{sub-$6$GHz}{Below $6\,$GHz}
\newacronym{ncr}{NCR}{Network-Controlled Repeater}
\newacronym{nlos}{NLoS}{Non-Line of Sight}
\newacronym{src}{SRC}{Smart Radio Connection}
\newacronym{srd}{SRD}{Smart Radio Device}
\newacronym{cs}{CS}{Candidate Site}
\newacronym{tp}{TP}{Test Point}
\newacronym{fov}{FoV}{Field of View}
\newacronym{nrric}{near-RT RIC}{Near Real-time {RAN} Intelligent Controller}
\newacronym{e2ap}{E2AP}{E2 Application Protocol}
\newacronym{e2sm}{E2SM}{E2 Service Model}
\newacronym{nrtric}{non-RT RIC}{Non-Real-Time {RIC}}
\newacronym{itti}{ITTI}{Inter-task Interface}
\newacronym{bap}{BAP}{Backhaul Adaptation Protocol}
\newacronym{iabest}{IABEST}{Integrated Access and Backhaul Experimental large-Scale Tetbed}
\newacronym{teid}{TEID}{Tunnel Endpoint Identifier}
\newacronym{dlsch}{DL-SCH}{Downlink Shared Channel }
\newacronym{ulsch}{UL-SCH}{Uplink Shared Channel }
\newacronym{rsu}{RSU}{Road Side Unit}
\newacronym{its}{ITS}{Intelligent Transportation Systems}
\newacronym{vanet}{VANET}{Vehicular Ad-hoc Network}
\newacronym{dt}{DT}{Digital Twin}
\newacronym{ecc}{ECC}{Edge Computing Cluster}
\newacronym{fig}{Fig.}{Figure}
\newacronym{wab}{WAB}{Wireless Access Backhaul}
\newacronym{cpe}{CPE}{Customer Premises
Equipment}
\newacronym{ntn}{NTN}{Non-Terrestrial Network}
\newacronym{mwab}{MWAB}{Mobile Wireless Access Backhaul}
\newacronym{tn}{TN}{Terrestrial Network}
\newacronym{eirp}{EIRP}{Effective Isotropic Radiated Power}
\newacronym{mbs}{MBS}{Mobile Base Station}
\newacronym{mbsr}{MBSR}{Mobile Base Station Relay}

\usepackage{fancyhdr}
\fancypagestyle{firstpage}{
  \fancyhf{}
  
  \fancyhead[C]{\small Accepted to IEEE Vehicular Networking Conference (VNC) 2026. \\ Please cite as: C. Rubaltelli, M. Morini, E. Moro, and I. Filippini, ``Enabling Mobile Base Stations in 5G via Wireless Access Backhaul (WAB),'' \emph{IEEE Vehicular Networking Conference (VNC)}, 2026.}
}

\begin{document}



\title{Enabling Mobile Base Stations in 5G via Wireless Access Backhaul (WAB): A Multi-Band Experimental Study}

\author{\IEEEauthorblockN{Chiara Rubaltelli, Marcello Morini, Eugenio Moro, Ilario Filippini}
\IEEEauthorblockA{Dipartimento di Elettronica, Informazione e Bioingegneria, 
Politecnico di Milano, Milan, Italy\\
$\langle name \rangle$.$\langle surname \rangle$@polimi.it}

\vspace{-8mm}
}

\maketitle
\thispagestyle{firstpage}

\begin{abstract}
Highly dynamic and mobile applications, such as vehicular networks, require stable connectivity, which is often challenging to achieve. Network densification is a key approach to address this issue and can be achieved cost-effectively through mobile base stations and wireless relaying. However, existing solutions rely on rigid and complex architectures that hinder deployment in dynamic scenarios.
The recently standardized \gls{wab} architecture represents a key evolution, enabling flexible and modular wireless relay networks with native support for mobility and multi-technology wireless backhaul.
This paper presents the first experimental realization of a multi-band \gls{wab} testbed, combining an FR2 backhaul and an FR1 access link using open-source software and commercial off-the-shelf components. The proposed framework validates end-to-end \gls{wab} operation under mobility and demonstrates the extension of FR2 coverage while maintaining compatibility with legacy FR1 user equipment.
Experimental campaigns in vehicular and outdoor-to-indoor scenarios confirm that \gls{wab} effectively mitigates FR2 limitations, particularly in uplink and Non-Line-of-Sight conditions. These results highlight \gls{wab} as a practical and scalable approach for vehicular and next-generation wireless networks.

\label{sec:abstract}
\end{abstract}


\glsresetall

\section{Introduction}
The growing demand for bandwidth and spectral efficiency in wireless networks has intensified research into new frequency bands and novel \gls{ran} architectures for \acrshort{5g}. Network \textit{densification} is a key enabler to sustain this continuous growth in radio access capacity demand, particularly in urban vehicular environments. Such environments are characterized by high user mobility, rapidly varying traffic conditions, and bandwidth-intensive applications such as autonomous driving, which require high-data-rate video streams.
Densification also facilitates the adoption of higher frequency bands for radio access, such as Frequency Range~2 (FR2, 24.25--71~GHz), which offers wide bandwidths and supports multi-Gbps data rates, but suffers from severe path loss, high penetration losses, and strong susceptibility to environmental dynamics. These effects are particularly pronounced in vehicular scenarios~\cite{mmwave}.

To mitigate the deployment costs associated with dense infrastructure in vehicular networks, \textit{wireless relaying} provides a cost-effective means to increase the number of \glspl{bs}. Furthermore, \glspl{mbs} represent a promising solution to address the strong temporal and spatial variability of traffic demand in vehicular environments and to explore further cost–capacity trade-offs. Indeed, as shown in~\cite{mobile_base_stations}, substantial cost savings can be achieved by deploying \glspl{mbs} in place of static ones. Consequently, a mobility-capable radio access infrastructure based on wireless relaying becomes particularly attractive for vehicular and dynamic urban scenarios, where rapid reconfiguration and service continuity under mobility are essential.

To enable effective wireless relaying, \gls{3gpp} took an initial step in 2019 by introducing \gls{iab} in Release~15, based on the wireless backhaul paradigm. Several enhancements have been proposed in subsequent releases to improve the technology; however, \gls{iab} has not achieved widespread adoption yet. The intrinsic complexity of the \gls{iab} architecture, together with its limited flexibility, makes it poorly suited to highly dynamic scenarios, such as vehicular networks and mobile base station deployments, which demand frequent backhaul reconfiguration and seamless backhaul handovers under mobility.

For these reasons, the Mobile Base Station Relay (MBSR) topic has quickly gained momentum within \gls{iab} studies, resulting in the introduction and standardization of \gls{wab} in Release~19~\cite{bridge_6G}. \gls{wab} leverages the benefits of wireless backhaul while relying on standard 5G components and procedures, thereby enabling a rapid and cost-efficient deployment.
Each \gls{wab} node integrates a \gls{mt} with a complete \gls{gnb}, enabling direct access to \glspl{ue} and seamless backhaul handovers.

Furthermore, the modular design of \gls{wab} supports multiple backhaul technologies, including \glspl{ntn}, and introduces full independence between the access and backhaul segments. This feature enables \gls{wab} deployments in highly dynamic environments, such as trains, ships, and airplanes, where terrestrial backhaul availability is limited or discontinuous.

In this context, \gls{wab} represents a promising evolution in the field of \glspl{mbs}. However, to date, no physical implementation has been publicly documented. This paper addresses this gap by presenting a multi-band \gls{wab} testbed that integrates a commercial backhaul with an access segment using open-source software, \glspl{sdr}, and commercial components. Leveraging the multi-technology capabilities of \gls{wab}, the proposed solution combines the high throughput offered by FR2 communications in the backhaul with the wider device compatibility and superior penetration of Frequency Range 1 (FR1, 410--7125~MHz) in the access segment.
The proposed testbed, assembled from \gls{cots} components, demonstrates that the implementation of a \gls{wab} architecture can be achieved rapidly and cost-effectively using existing technological expertise. These results suggest that the widespread adoption of \gls{wab} may be closer to realization than previously anticipated.

\begin{figure*}[ht!]
    \centering
    \includegraphics[width=0.7\textwidth]{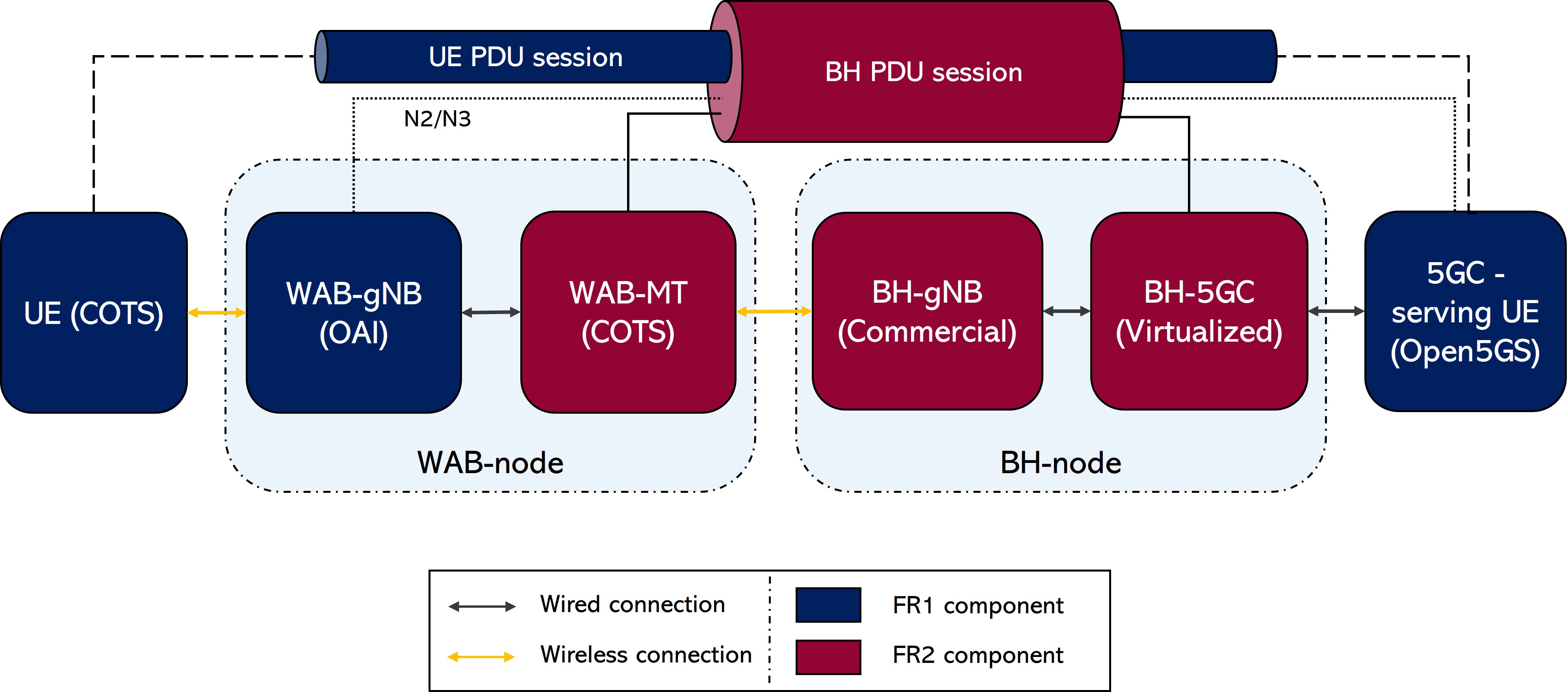}
    \caption{WAB's fundamental architecture, including the corresponding elements in the testbed.\vspace{-5mm}}
    \label{fig:wab_scheme}
\end{figure*} 

Using this testbed, we conducted an experimental campaign to validate the proposed system in urban vehicular and \gls{o2i} scenarios, confirming its potential to extend the benefits of FR2 communications by mitigating typical FR2 propagation impairments.
This initial campaign highlights the main advantages of wireless backhaul. In particular, the results demonstrate that \gls{wab} nodes can maintain robust connectivity under mobility. Moreover, the \gls{wab} architecture effectively overcomes uplink transmission limitations, providing enhanced performance even in challenging \gls{nlos} indoor environments.

The main contributions of this work can be summarized as follows:
\begin{itemize}
    \item We provide an overview and discussion of the \gls{wab} architecture as a key enabler for mobility-aware and vehicular mobile base station deployments.
    \item We present, to the best of our knowledge, the first physical implementation of the \gls{wab} architecture, realizing a complete multi-band testbed that combines a commercial FR2 wireless backhaul with an FR1 access segment based on open-source software.
    \item We experimentally validate the proposed system in realistic urban vehicular and \gls{o2i} scenarios, demonstrating robust connectivity under mobility and improved performance in challenging propagation conditions.
\end{itemize}

The remainder of this paper is organized as follows.
Section~\ref{sec:background} reviews the related work.
Section~\ref{sec:wab} outlines the \gls{wab} architecture and its fundamental operations. Section~\ref{sec:testbed} details the proposed multi-band testbed and its main components. Section~\ref{sec:validation} presents the experimental campaign and discusses the obtained results. Finally, Section~\ref{sec:conclusion} concludes the paper and highlights future research directions.

\label{sec:introduction}

\section{Background on Wireless Backhaul}
\label{sec:background}
Prior to \gls{wab}, \gls{3gpp} standardized several wireless relay architectures. Introduced in Release~15~\cite{iab_5g_mmw}, \gls{iab} enables wireless backhaul and supports multi-hop topologies, in which a donor node -- the only node with a direct connection to the \gls{5gc} -- provides backhaul connectivity to \gls{iab} nodes. These nodes can, in turn, act as parent nodes for other \gls{iab} nodes or directly serve \glspl{ue}.
Relying on a disaggregated \gls{ran} architecture, each \gls{iab} node comprises a \gls{mt}, used to connect upstream to the donor or parent node over the backhaul link, and a \gls{du}, responsible for providing access to \glspl{ue} or downstream child nodes. The \gls{cu} is located at the donor node.

Despite its conceptual advantages, \gls{iab} faces significant implementation complexity and limited flexibility, which have hindered its large-scale adoption, particularly in highly dynamic scenarios such as vehicular networks. Specifically, mobility procedures are rigid and verbose, additional layers are introduced in the 5G protocol stack, and backhaul handovers, although supported by the standard, remain extremely complex to realize in practice~\cite{miab}.

To address some of the challenges of \gls{iab}, \gls{3gpp} introduced \gls{mbsr}~\cite{shaping_6G}, an evolution of \gls{iab} designed to support more flexible mobility by adding a dedicated mobile \gls{cu} to assist \gls{iab} nodes when changing their donor. However, \gls{mbsr} still relies on centralized, full-network control, with resource management tightly coupled to that of the donor, as in conventional \gls{iab} architectures.
As a result, system complexity remains high, and scalability issues become increasingly pronounced as the number of backhaul hops grows.

In addition to architectural limitations, practical \gls{iab} implementations remain scarce. Although several analytical and simulation-based studies have investigated different aspects of \gls{iab}~\cite{e2e_iab, iabest}, experimental frameworks are very limited~\cite{miabjapan, uaviab}. Moreover, to the best of our knowledge, \gls{iab} nodes are currently available only in the form of experimental or pre-commercial products. The combined architectural complexity and lack of real-world validation appear to be the main obstacle to widespread \gls{iab} market deployment.

To overcome these limitations, \gls{3gpp} introduced \gls{wab} in Release~19. In 2025, \gls{wab} was standardized as a topological enhancement for 5G-Advanced (5G-A)~\cite{wabtechreport38799}.
In parallel, \gls{mwab} was standardized as an architectural enhancement targeting Vehicle-Mounted Relays (VMRs)~\cite{wabtechreport2370006}. Although defined by different Working Groups (WGs), the two technologies share the same architectural foundation, which is described in the next section.

Despite its strong potential, the existing literature on \gls{wab} remains limited. Current works primarily discuss \gls{wab} as a conceptual enabler for \gls{ntn} integration in 5G and 5G-Advanced networks~\cite{6g_connected_sky, 6g_ntn_tn}, without providing physical implementations or experimental validation. To the best of our knowledge, this work presents the first practical multi-band \gls{wab} testbed, together with its experimental validation in vehicular and \gls{o2i} scenarios.

\section{Wireless Access Backhaul}
\label{sec:wab}

This section presents the \gls{wab} architecture and its fundamental procedures, followed by an overview of representative application scenarios that highlight its applicability to highly dynamic mobile environments.

\subsection{Architecture}
The \gls{wab} architecture comprises two communication systems: one for radio access and one for wireless backhaul. These systems can operate in either in-band or out-of-band configurations, with the objective of providing end-to-end connectivity through wireless backhaul. A schematic representation of the basic architecture is shown in Fig.~\ref{fig:wab_scheme}.

One of the core elements of the \gls{wab} architecture is the \gls{wab} Node, which consists of a WAB-gNB and a WAB Mobile Termination (WAB-MT). The WAB-gNB is a standard-compliant 5G \gls{bs} that provides radio access to \glspl{ue}, while the WAB-MT supports a subset of \gls{ue} functionalities and establishes a wireless connection with the BH-gNB. This connection is used by the WAB-gNB as a backhaul link toward the \gls{cn}, referred to as 5GC-Serving-UE.  

Each \gls{wab} Node is associated with a donor node, referred to as the BH node, which includes the BH-gNB and a BH-5GC. In the basic configuration, the BH-gNB is also a 5G standard-compliant \gls{bs}, that provides connectivity to the WAB-MT through conventional radio access procedures and resources. The BH-gNB is locally connected to the BH-5GC, whose N6 interface links to the 5GC-Serving-UE. Alternatively, the backhaul segment can be implemented using a non-\gls{3gpp} link or network.  

In all cases, end \glspl{ue} remain fully agnostic to the underlying backhaul mechanisms, maintaining full service transparency. This property represents one of the key flexibility features of \gls{wab}, enabling the extension of 5G coverage through heterogeneous and non-\gls{3gpp} infrastructures.

The \gls{wab} architecture enables end-to-end connectivity for the end \gls{ue} by tunneling the N2 and N3 interfaces -- linking the WAB-gNB to the 5GC-Serving-UE -- over Backhaul (BH) \gls{pdu} sessions established between the WAB-MT and the BH-5GC.  
In the control plane, the wireless backhaul is provided through a BH PDU session that carries the N2 interface between the WAB-gNB and the \gls{amf} of the 5GC-Serving-UE. Similarly, in the user plane, traffic is conveyed over a separate BH PDU session supporting the N3 interface between the WAB-gNB and the \gls{upf} of the 5GC-Serving-UE.

The backhaul (red in Fig.~\ref{fig:wab_scheme}) and access (blue in Fig.~\ref{fig:wab_scheme}) segments operate as two logically independent networks. The backhaul network provides backhaul connectivity to the WAB-MT, while the access network serves the end \gls{ue}. The backhaul network can be implemented over either a \gls{tn} or a \gls{ntn}; in both cases, 5G access services remain transparent to end \glspl{ue}, regardless of the underlying backhaul technology.

\subsection{Basic Procedures}
The WAB-node attachment, also referred to as WAB-node integration, begins with the setup of the WAB-MT, which connects to the BH-gNB as a standard \gls{ue} using conventional 5G access procedures. Once authorized by the BH-5GC, the WAB-MT establishes one or more BH PDU sessions with the \gls{upf} of the BH-5GC. Multiple BH PDU sessions can be created to transport different logical interfaces or to enable differentiated traffic handling based on \gls{qos} requirements. These sessions form the foundation for carrying traffic from the WAB-gNB and its connected \glspl{ue}.

Following the WAB-MT setup, the WAB-gNB is initialized and registers with the 5GC-Serving-UE, thereby enabling the establishment of PDU sessions for connected \glspl{ue}. Specifically, each UE PDU session is supported by a Data Radio Bearer (DRB) between the UE and the WAB-gNB, and by a conventional N3 \gls{gtp} tunnel between the WAB-gNB and the 5GC-Serving-UE. The N3 tunnel is encapsulated within the BH PDU session tunnels established during the WAB-node integration phase.

Through the Xn interface, the WAB-gNB can establish connections with neighboring gNBs. These connections are tunneled through the backhaul network and may terminate either at the BH-gNB serving the WAB-MT or at other \glspl{gnb} connected to the 5GC-Serving-UE. 

As the \gls{wab} node moves, the WAB-MT follows legacy \gls{ue} mobility procedures. This behavior represents one of the key mobility features of \gls{wab}, enabling seamless handovers within the backhaul network. Moreover, the WAB-MT can request the establishment or modification of BH PDU sessions based on the traffic requirements of the WAB-gNB as it moves across the network. To prevent multi-hop across mobile \gls{wab} nodes, a WAB-MT is not allowed to hand over to a target WAB-gNB. As a result, star-like, single-hop logical topologies are formed, interconnecting mobile \gls{wab} nodes with the donor node. However, this single-hop logical constraint does not restrict the backhaul network itself, which may employ arbitrary underlying technologies or topologies to support the user- and control-plane tunneling required by mobile \gls{wab} nodes.

Free mobility and seamless backhaul handovers are the two key features that distinguish the \gls{wab} architecture from similar solutions, such as NR femtocells, which are primarily designed to provide NR access in residential or enterprise environments. 
An important research direction for WAB adoption in mobile and vehicular scenarios is the design of efficient interference management and coordination techniques, as mobile WAB nodes may generate interference when operating within static infrastructures due to imperfect synchronization.

\vspace{-0.2cm}
\subsection{Mobility Applications}
Thanks to seamless mobility support and its multi-technology backhaul design, \gls{wab} enables highly dynamic urban applications. 
A \gls{wab} node can carry a local 5G access \emph{bubble}, allowing nearby users and vehicles to connect as the node moves through the environment. This capability can be leveraged to provide 5G connectivity on trains or other long-distance transportation systems, where coverage is often intermittent. By deploying a \gls{wab} node on board, only the WAB-MT performs handovers when the train moves, while a large number of \glspl{ue} remain attached to the same serving \gls{gnb}. 
This significantly reduces the signaling overhead and core network load, while improving service quality and continuity, even across tunnels or national borders~\cite{moving_net}.

Similarly, \gls{wab} systems can support temporary or emergency deployments, such as disaster recovery scenarios or large-scale public events (e.g., earthquakes, floods, or mobile crowds and city-wide sport events). In these cases, mobile \gls{wab} nodes, potentially obtaining backhaul connectivity via \gls{ntn}, can rapidly extend coverage in areas where traditional terrestrial infrastructures are unavailable or highly congested.

\section{WAB Experimental Framework}
\label{sec:testbed}
The realized testbed aligns with the fundamental architecture of \gls{wab} segments and the corresponding components are displayed in Fig. \ref{fig:wab_scheme} (in brackets). The experimental setup implements a multi-band \gls{wab} configuration, where the access segment -- connecting \gls{ue} and WAB-gNB -- operates at FR1, while the backhaul segment -- connecting WAB-MT and BH-gNB -- works at FR2.

A detailed description of the testbed components is provided below, highlighting how they implement a \gls{3gpp}-compliant, fully functional \gls{wab} system using open-source software and commercial hardware, ensuring compatibility with \gls{cots} \glspl{ue}. The assembly required no hardware modifications beyond standard 5G components. 
This simplicity is key to achieve rapid and cost-effective deployment, thereby facilitating the widespread adoption of \gls{wab} technology.

\subsection{FR2 components}
The FR2 segment components are part of the High-Frequency Campus Lab (HFCL)~\cite{hfcl}, an open and standard-compliant FR2 5G network deployed at Politecnico di Milano. Specifically, the WAB-MT is implemented using a \gls{cots} \gls{cpe}.  
The BH-gNB's \gls{aau} is installed at a height of 22~m on a building rooftop and is interfaced via a 25~Gbps eCPRI fiber fronthaul to a \gls{bbu}. The \gls{bbu} is connected to a virtualized commercial \gls{5gc} (BH-5GC) through a 10~Gbps backhaul link.

The segment operates at a center frequency of 27.2~GHz, with a channel bandwidth of 200~MHz and a subcarrier spacing of 120~kHz. The frame structure follows a \gls{tdd} 4:1 configuration. The \gls{aau} provides an \gls{eirp} of 70~dBm with an 8T8R \gls{mimo} configuration, while the \gls{cpe} operates with an \gls{eirp} of 40~dBm and a 2T2R \gls{mimo} configuration.  

Further details on the FR2 segment are listed in Table~\ref{table:param}. The FR2 network is used exclusively for research purposes and was entirely dedicated to the testbed, without any additional traffic.

\begin{table}[t]
    \centering 
    \caption{Operational parameters of BH FR2 components.}
    \begin{tabular}{ c | c }
    \hline
    \textbf{Parameter} & \textbf{Value}   \\
    \hline \hline
    \textbf{AAU coordinates (lat, lon, h)} & 45.478671, 9.232550, 22 m \\
    \textbf{AAU azimuth, down tilt} & 135°, 2° \\
    \textbf{Center frequency} & 27.2 GHz \\
    \textbf{Channel bandwidth} & 200 MHz \\
    \textbf{Subcarrier spacing} & 120 kHz \\
    \textbf{Frame structure} & TDD 4:1 \\
    \textbf{Maximum QAM order (D/U)} & 256/64 \\
    \textbf{AAU TX power} & 37.5 dBm \\
    \textbf{AAU gain} & 32.5 dBm \\
    \textbf{AAU EIRP} & 70 dBm \\
    \textbf{AAU MIMO} & 8T-8R \\
    \textbf{CPE gain} & 20 dBi \\
    \textbf{CPE EIRP} & 40 dBm \\
    \textbf{CPE MIMO} & 2T-2R \\
    \hline
    \end{tabular}
    \vspace{-3mm}
    \label{table:param}
\end{table}

\begin{figure*}[!ht]
    \centering

    \includegraphics[width=1.7\columnwidth]{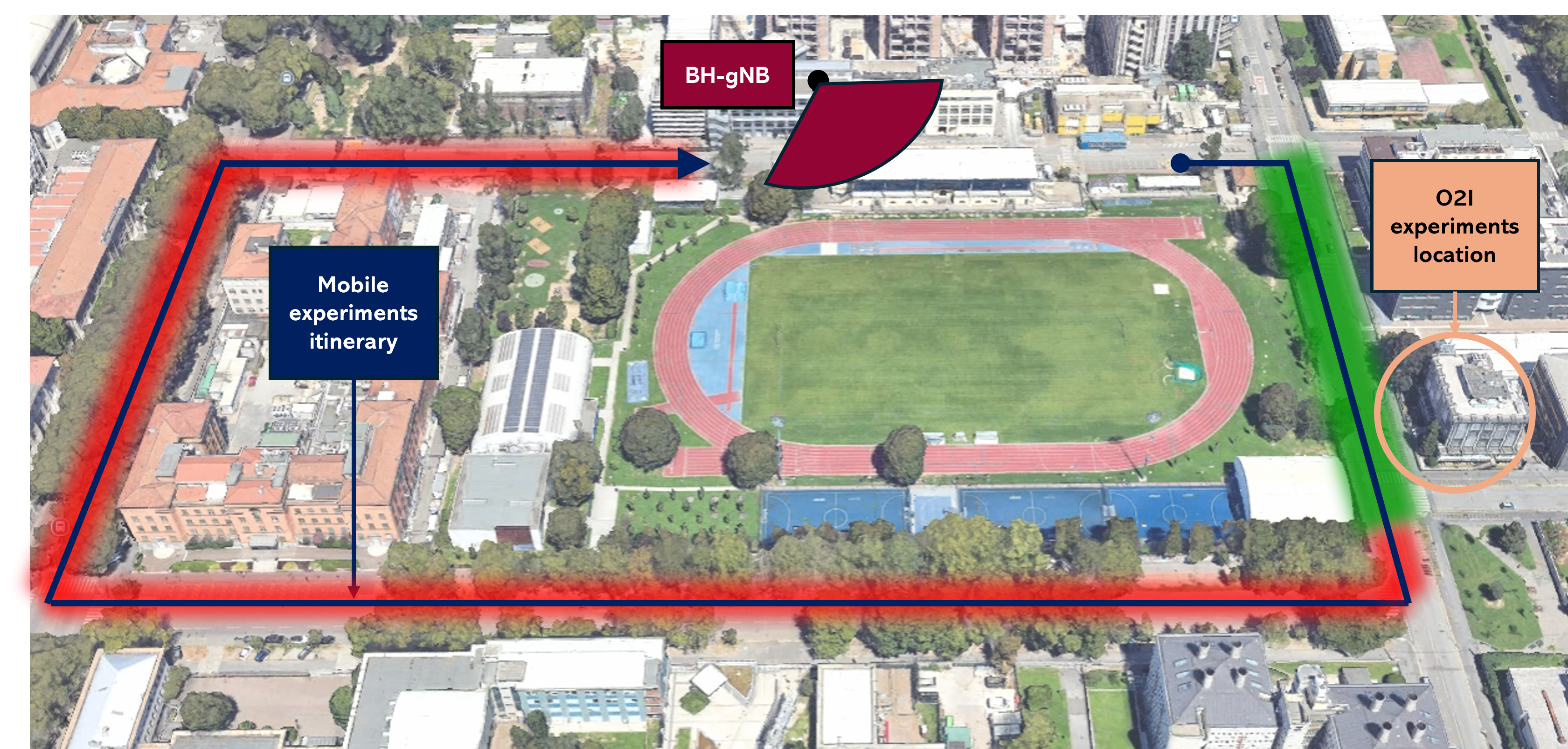}
    \caption{Aerial view of the urban neighborhood covered by the FR2 \gls{bs} together with experiment locations. Mobile experiment trajectory shows \gls{nlos} and \gls{los} regions highlighted, respectively, in red and green.\vspace{-5mm}
    }
    \label{fig:satellite}
\end{figure*}

\subsection{FR1 components}
The \gls{ue} used in the testbed is a \gls{cots} 5G smartphone operating in FR1. The WAB-gNB is implemented using an \gls{oai} gNB, deployed on a laptop equipped with an Ettus \gls{usrp} B210 \gls{sdr}. \Gls{oai} is an open-source platform that enables the deployment of a 5G \gls{ran} on general-purpose hardware, which can be connected to \glspl{sdr} to provide wireless access to commercial \glspl{ue}~\cite{openairinterface}.

The FR1 gNB operates with 40~MHz bandwidth and a subcarrier spacing of 30~kHz in a SISO configuration. The maximum transmission power is 20~dBm, and the \gls{tdd} configuration follows a 4:1 downlink-to-uplink split. The WAB-gNB laptop is connected to the WAB-MT via an Ethernet link.  
Finally, the 5GC-Serving-UE consists of an Open5GS Core instance installed on a general-purpose server connected to the BH-5GC. Open5GS is an open-source implementation of the \gls{5gc}, composed of virtualized Network Functions.

\subsection{Connectivity}
Once all components were assembled into the final \gls{wab} testbed, connectivity was established as follows. Since the 5GC-Serving-UE could not be directly reached by the WAB-gNB due to the \gls{cpe} \gls{nat} mechanism, a Wireguard tunnel was deployed to transport the N2 and N3 interfaces between the 5GC-serving-UE and the WAB-gNB. Wireguard tunnels are open-source Virtual Private Networks (VPNs) that utilize state-of-the-art cryptographic protocols, offering a secure and low-overhead solution. This configuration reflects real-world deployments, where the N2 and N3 interfaces are typically protected from direct exposure to external networks. Some adjustments to the tunnel configuration were required to optimize performance. In particular, the \gls{mtu} size was reduced from 1420~bytes to 1384~bytes, as suggested in~\cite{mtugithub}.

In summary, the proposed setup employs three levels of encapsulation to ensure proper end-to-end connectivity: the BH PDU tunnel encapsulates the VPN tunnel transporting the N2 and N3 interfaces, which in turn carries the \gls{pdu} sessions of the end \glspl{ue}.

\section{Validation and Results}
\label{sec:validation}




Dense urban environments pose significant challenges to high-frequency signals in the FR2 band, where buildings and other obstacles cause severe penetration and propagation losses. Multi-band \gls{wab} technology mitigates these effects by jointly exploiting FR2 and FR1 links to extend coverage and improve reliability. In particular, an FR1 hop can overcome blockages that would otherwise disrupt FR2-only connectivity.
Moreover, while all \glspl{ue} support FR1 operation, only a limited subset can directly access FR2 \glspl{ran}. A mobile \gls{wab} system enables transported and nearby FR1 \glspl{ue} to benefit from the high capacity provided by FR2 backhaul connectivity.

To validate the effectiveness of \gls{wab} under these conditions, we conducted an extensive experimental campaign focused on the operation of a mobile \gls{wab} node in a dense urban environment. A fully mobile \gls{wab} node was assembled and used to perform measurements while moving through the urban area, within the service region of a single FR2 sector providing backhaul connectivity. In addition, we carried out \gls{o2i} experiments to complement the mobility evaluation and assess system performance under static yet propagation-challenging conditions.

Figure~\ref{fig:satellite} shows an aerial view of the urban neighborhood where the tests were performed, indicating the placement and orientation of the BH-gNB.  
In both scenarios, the BH-5GC and 5GC-Serving-UE were located inside the building hosting the BH-gNB, while the WAB node and end \gls{ue} were moved around the area. The \gls{cpe} (WAB-MT) was powered by a power bank and connected to a laptop running probe software to collect FR2 link measurements, and via Ethernet to the general-purpose machine running the \gls{oai} gNB for FR1 log acquisition through \gls{oai} logging functions.  
A Secure Shell (SSH) tunnel between this machine and the 5GC-Serving-UE server enabled remote execution of \textit{iperf3} commands on both the server and the \gls{ue}. Speed tests were conducted using \gls{tcp} connections to evaluate end-to-end throughput.

\begin{figure}[!h]
    \centering
    \includegraphics[width=1\columnwidth]{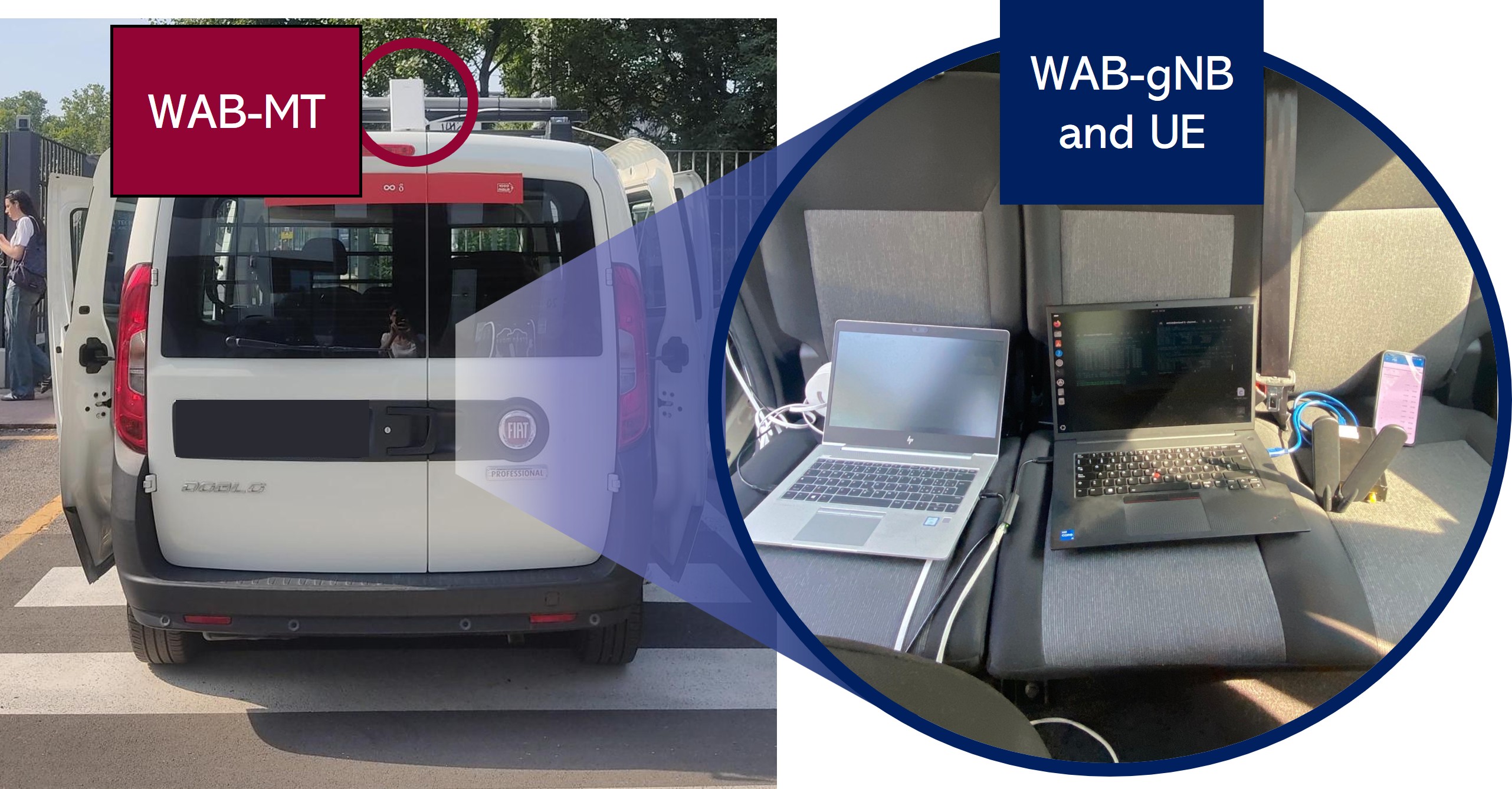}
    \caption{{Mobile \gls{wab} node.}}
    \label{fig:mwab_node}
\end{figure}

\subsection{Mobile experiments}
In this scenario, the WAB node was installed on a vehicle, as shown in Figure \ref{fig:mwab_node}. The WAB-gNB and \gls{ue} were placed inside the cabin, while the WAB-MT (FR2 \gls{cpe}) was mounted on the vehicle rooftop. The vehicle moved at up to 30~km/h.  
Given the high directivity of the \gls{cpe} antenna array and possible signal losses or reflections from the vehicle body, the WAB-MT was mounted at the rear, with its antenna array oriented perpendicular to the ground. This setup ensured optimal reception when traveling south in the \gls{los} area. A GPS device tracked the route and georeferenced the measurements.  
The chosen itinerary looped around an outdoor sports center and included segments with trees and buildings causing partial signal obstruction. End-to-end throughput measurements were performed in both \gls{dl} and \gls{ul} directions.

\begin{figure}[!h]
    \centering
    \begin{subfigure}[b]{0.45\textwidth}
        \centering
        \includegraphics[width=\linewidth]{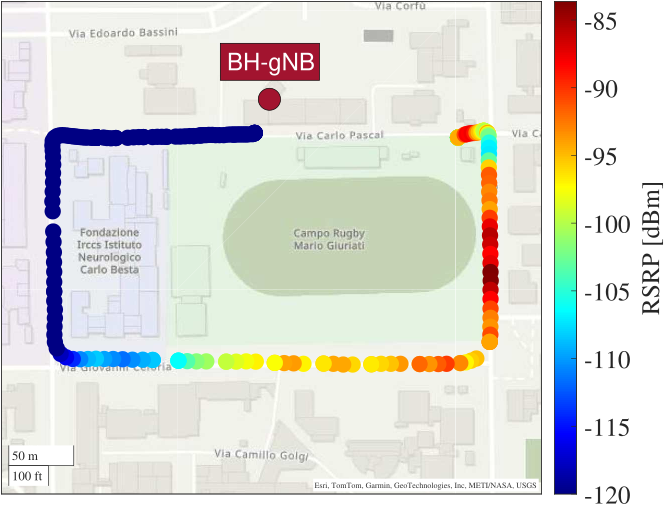}
        \caption{\gls{dl} FR2 RSRP map}
        \label{fig:rsrp}
    \end{subfigure}
    \hfill
    \begin{subfigure}[b]{0.45\textwidth}
        \centering
        \includegraphics[width=\linewidth]{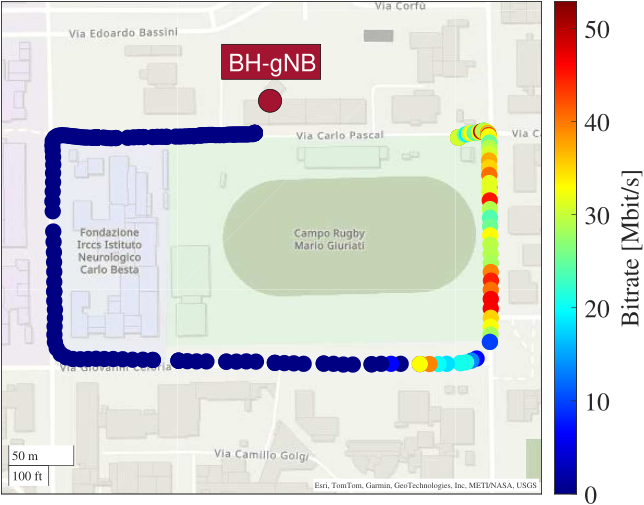}
        \caption{\gls{dl} end-to-end throughput map}
        \label{fig:br}
    \end{subfigure}
    \caption{Mobile \gls{wab} measurements of FR2 RSRP and end-to-end throughput. }
    \label{fig:results}
    \vspace{-5mm}
\end{figure}

\begin{figure}[!h]
     \centering
     \begin{subfigure}[]{0.48\textwidth}
         \centering
         \includegraphics[width=\textwidth]{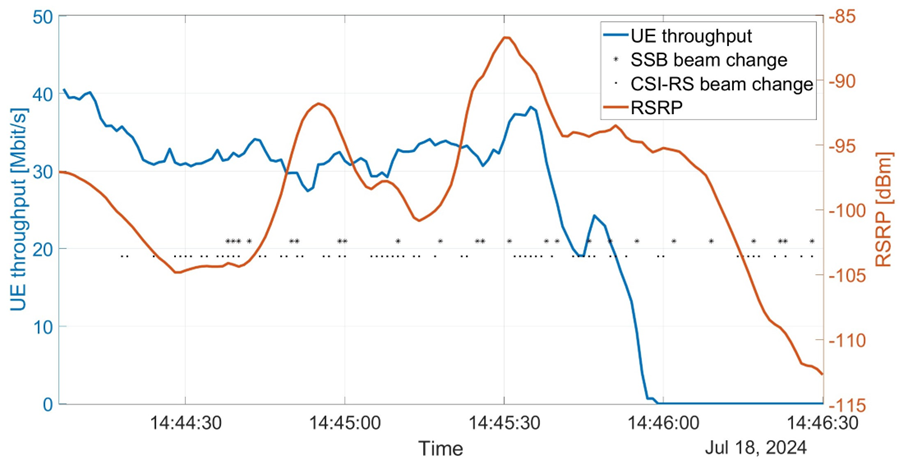}
         \centering
         \caption{End-to-end throughput, FR2 \gls{rsrp} and beam changes in mobile \gls{wab} experiments.}
         \label{fig:rsrp_time}
     \end{subfigure}
     \hfill
     \begin{subfigure}[]{0.48\textwidth}
         \centering
         \includegraphics[width=\textwidth]{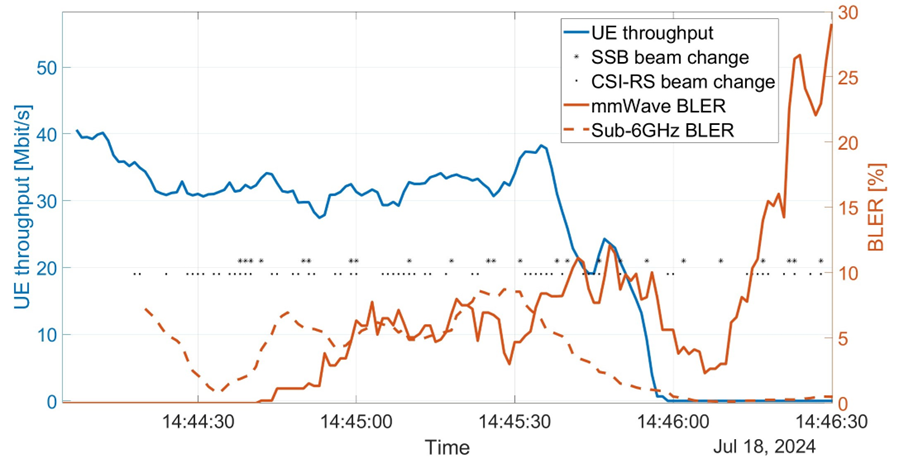}
         \caption{End-to-end throughput, FR2 and FR1 \gls{bler} and beam changes in mobile \gls{wab} experiments.}
         \label{fig:bler_time}
     \end{subfigure}
        \caption{Mobile \gls{wab} experiment results plotted in time.\vspace{-0.2cm}}
        \label{fig:time_results}
\end{figure}

We first present the \gls{dl} results. The data show that the end-to-end throughput (i.e., UE throughput) strongly correlates with the FR2 backhaul link quality. As illustrated in Figure~\ref{fig:results}, the FR2 backhaul \gls{rsrp} peaks at --83~dBm in the \gls{los} zone, where the WAB-MT has direct visibility of the BH-gNB. In this region (until 14:45:40, as shown in Figure~\ref{fig:rsrp_time}), the end-to-end throughput reaches approximately 50~Mbps.  
As the vehicle moves into the \gls{nlos} area, performance gradually degrades: throughput begins to decline when a single building obstructs the signal and eventually drops to zero in deep \gls{nlos}. This throughput degradation occurs faster than the corresponding decrease in FR2 backhaul \gls{rsrp}.

This discrepancy is clarified in Figure~\ref{fig:bler_time}: as the end-to-end throughput decreases, the FR2 backhaul link exhibits an increasing \gls{bler}, explaining the abrupt drop in performance. The figure also reports the serving beams for both \gls{ssb} and \gls{csirs}. Under \gls{los} conditions, the wider \gls{ssb} beams remain stable, whereas in \gls{nlos} regions, beam switching begins. In contrast, the narrower and more channel-sensitive \gls{csirs} beams continuously adjust to maintain link quality. Beam switching, however, shows no observable impact on end-to-end throughput.

During the tests, the FR1 access link operated entirely inside the vehicle, where external factors had minimal influence. The \gls{mcs} remained stable around 27 for most of the experiment. In contrast, the FR2 segment faced more challenging conditions, with average \gls{mcs} values around 20, reflecting the effects of mobility, obstructions, and environmental dynamics on the backhaul link.

\Gls{ul} measurements exhibited a trend similar to the \gls{dl} results. End-to-end throughput peaked in the \gls{los} region, matching the maximum FR2 backhaul \gls{rsrp} values, and dropped rapidly as the vehicle entered \gls{nlos}. On average, the \gls{ul} throughput stabilized around 1~Mbps. The notable imbalance between end-to-end \gls{dl} and \gls{ul} throughput primarily arises from the configuration of the mmWave backhaul segment. This setup is unfavorable for the \gls{ul} direction: the 4:1 \gls{tdd} frame prioritizes \gls{dl} traffic, the \gls{cpe} transmits at lower power than the \gls{aau}, and the maximum supported \gls{qam} order is higher in \gls{dl}.  
Additional imbalance stems from the FR1 access segment, whose \gls{tdd} scheme mirrors that of the FR2 link.
 
The results from the vehicular experiments demonstrate that reliable FR2 coverage through the \gls{wab} architecture can be achieved in \gls{los} scenarios as well as in \gls{nlos} conditions where shadowing is not excessively severe. Moreover, the proposed architecture enables widely available \gls{cots} FR1 smartphones to access an FR2 \gls{ran} without requiring dedicated or costly FR2 radio modules. Due to the limited coverage of the available FR2 network, the testbed could not rely on multiple BH nodes; consequently, an experimental evaluation of backhaul handovers was left outside the scope of this study. Future work will focus on experimentally assessing backhaul handover performance under mobility.

\subsection{\gls{o2i} experiments}
\begin{figure}[!h]
    \centering
    \includegraphics[width=0.8\columnwidth]{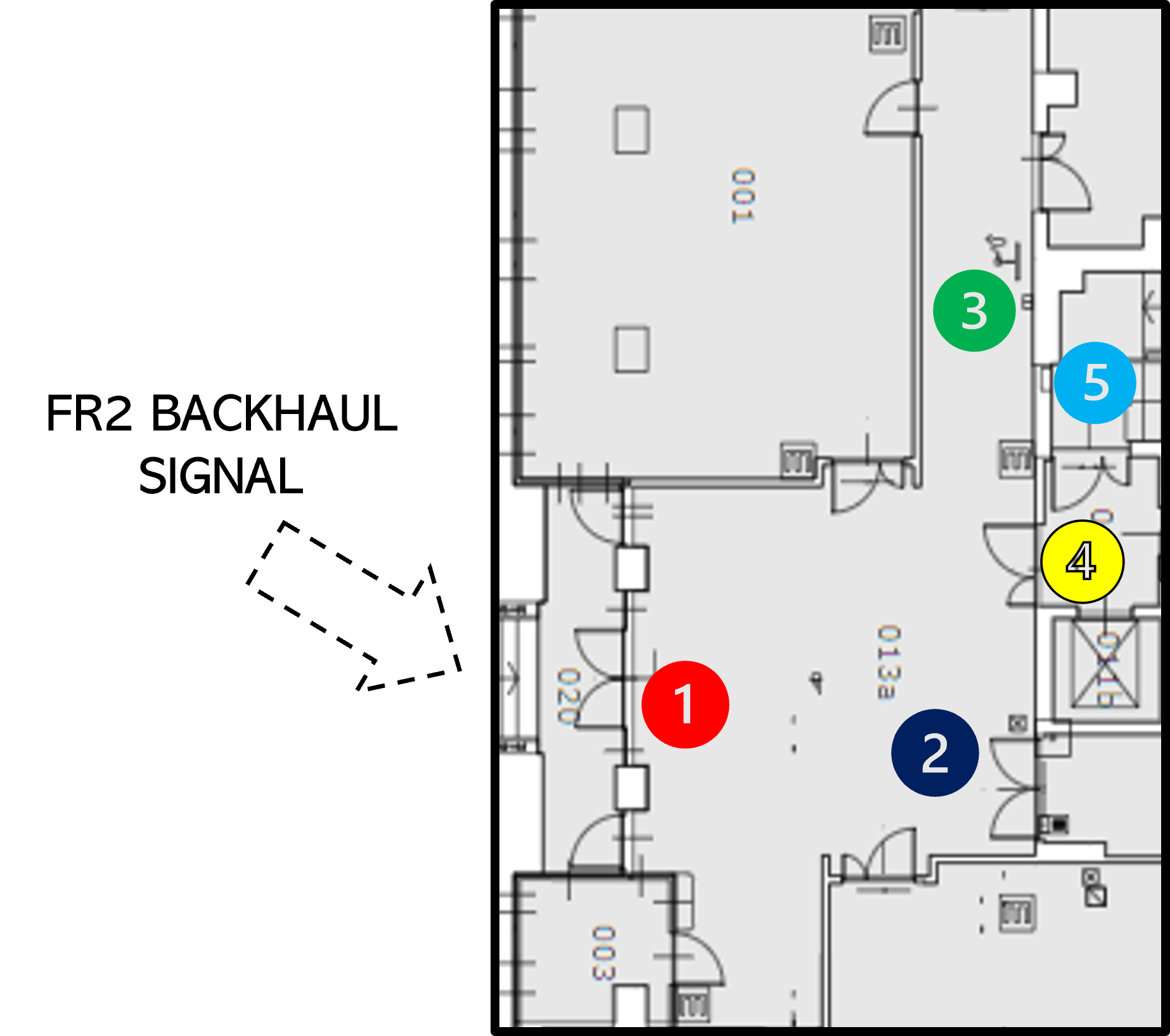}
    \caption{\gls{o2i} experiment positions.\vspace{-5mm} 
    }
    \label{fig:positions}
\end{figure}

In this scenario, the \gls{wab} node was placed inside a building with direct visibility of the BH-gNB.  
End-to-end throughput measurements were conducted at five indoor positions, as shown in Fig.~\ref{fig:positions}. Position~1 corresponds to a \gls{los} condition with the BH-gNB, with only a glass facade attenuating the signal. Positions~2 and~4 are located deeper inside the building but remain subject to glass attenuation only. Positions~3 and~5, situated behind interior walls, experience additional attenuation of the incoming signal.

We first conducted FR2-only speed tests by moving the \gls{cpe} through each of the five positions to assess the performance of the backhaul link. Subsequently, we assembled the \gls{wab} node and placed it at Position~1 to ensure the best possible signal conditions. We then conducted speed tests by moving the final \gls{ue}, connected to the \gls{wab} node via the FR1 link, through each of the five positions. Tests were performed in both the \gls{ul} and \gls{dl} directions.

Similarly to the mobile experiments, the directivity of the WAB-MT must be considered. During the FR2 measurements, both with and without the \gls{wab} node, the \gls{cpe} was oriented in the optimal direction, as recommended in~\cite{hfcl}, to ensure maximum throughput even under \gls{nlos} conditions. This reflects the reasonable assumption that any deployed antenna array would be installed with optimal alignment.

To enable a fair comparison between the 200~MHz FR2 CPE-only scenario and the 40~MHz FR1 WAB-gNB configuration, we computed the corresponding spectral efficiency results.  
However, it is important to note the substantial differences between the two implementations. The FR2 \gls{ran} is a commercial, optimized system running on dedicated hardware, operating over a 200~MHz bandwidth with a 2$\times$2 \gls{mimo} configuration (limited by the \gls{cpe}), a maximum transmit power of 37.5~dBm, and an antenna gain of 32.5~dBi.  
In contrast, the FR1 link is a prototype implementation based on virtualized functions running on a general-purpose machine with \glspl{sdr}, operating in SISO mode with a transmit power of only 20~dBm and a basic dipole antenna. Consequently, a significantly higher spectral efficiency is expected from the FR2 link compared to the FR1 one.

\begin{figure}[!h]
     \centering
     \begin{subfigure}[]{0.48\textwidth}
         \centering
         \includegraphics[width=\textwidth]{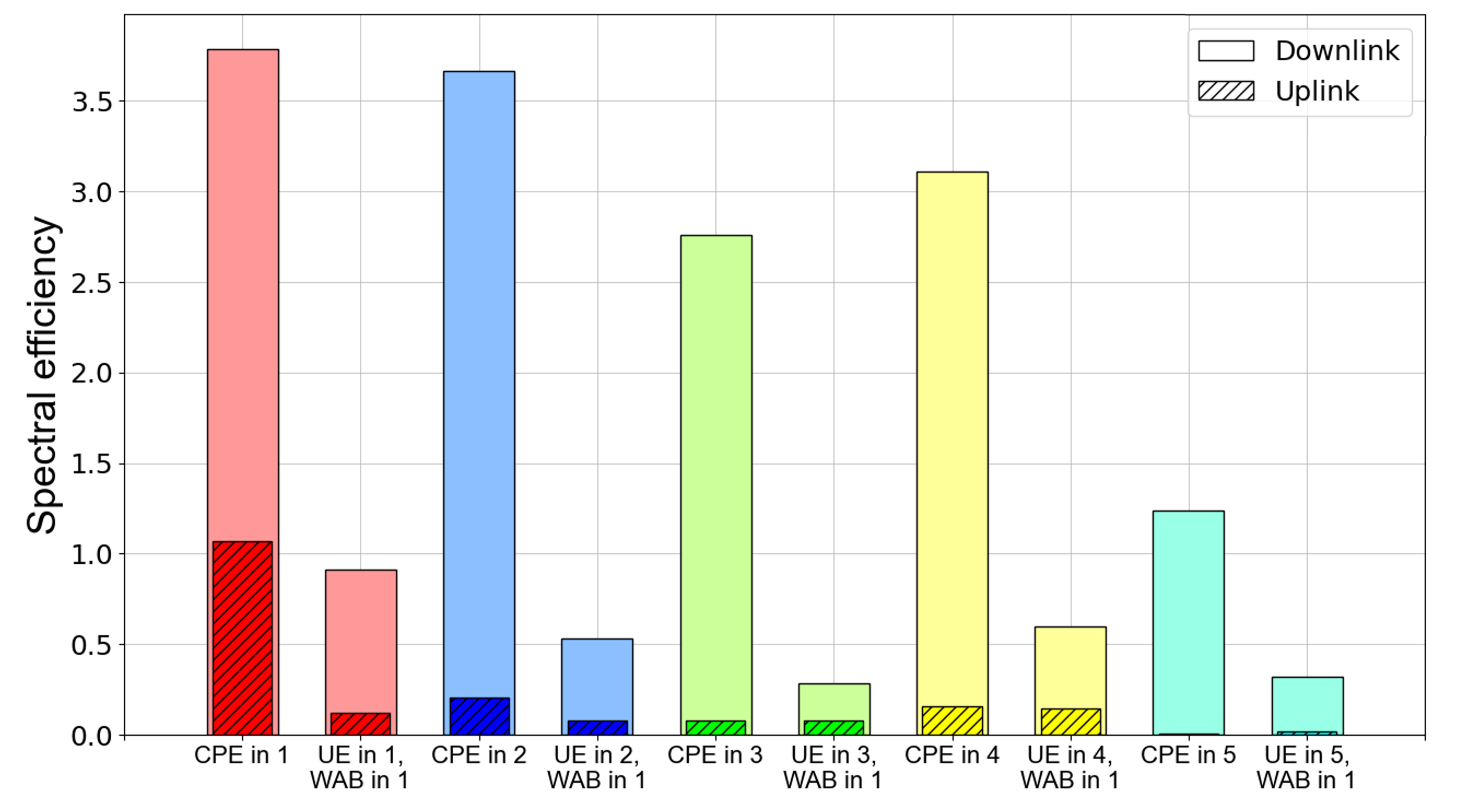}
         \centering
         \caption{\gls{dl} and \gls{ul} SEs for each measurements positions of O2I experiments. Bar colors correspond to positions colors used in Fig. \ref{fig:positions}.}
         \label{fig:barplot_1}
     \end{subfigure}
     \hfill
     \begin{subfigure}[]{0.48\textwidth}
         \centering
         \includegraphics[width=\textwidth]{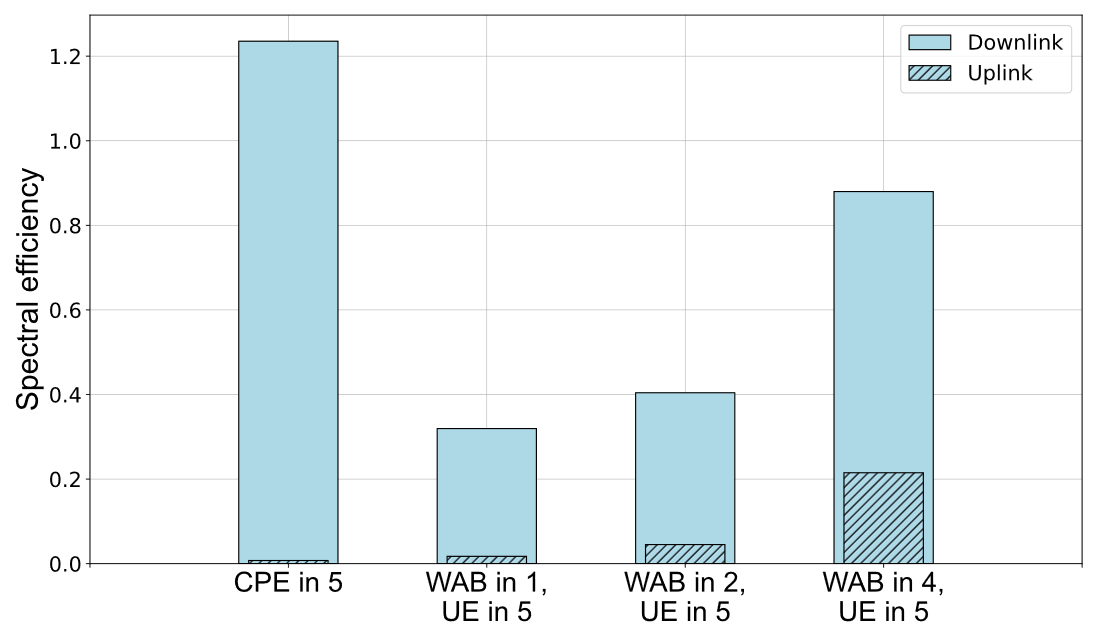}
        \caption{\gls{dl} and \gls{ul} SEs of the additional indoor campaign.}
         \label{fig:barplot_2}
     \end{subfigure}
        \caption{\gls{o2i} experiments spectral efficiencies.\vspace{-5mm}}
        \label{fig:barplots}
\end{figure}

The obtained results are shown in Fig.~\ref{fig:barplot_1}. In the \gls{dl} case, as expected, the FR2-only \gls{cpe} achieves significantly higher spectral efficiency than the \gls{wab} system, particularly at Positions~1 and~2, where the signal is attenuated only by glass.  
In the \gls{ul} case, FR2 performance degrades sharply when transitioning from \gls{los} to \gls{nlos} conditions. The \gls{wab} system also exhibits a performance drop between Position~1 and Positions~2 and~4, but the reduction is less pronounced. Notably, at Positions~3 and~5, where additional wall attenuation is present, the \gls{wab} system achieves comparable or even higher \gls{ul} spectral efficiency than the FR2-only \gls{cpe}.  
These results highlight the potential of the \gls{wab} architecture to mitigate FR2 limitations in challenging indoor \gls{nlos} environments, particularly in the uplink. Therefore, we further investigated Position~5, where the \gls{wab} system outperformed the FR2 \gls{cpe} in terms of \gls{ul} spectral efficiency.

In this additional campaign, the \gls{ue} was fixed at Position~5, while the \gls{wab} node was moved to Positions~2 and~4. This configuration maintained a stable backhaul link while reducing the distance between the \gls{wab} node and the \gls{ue}.  
This adjustment had a significant impact on the FR1 access link quality, given the limitations of our testbed due to the use of prototype \glspl{sdr}, the WAB-gNB operating with a low maximum transmit power, and its signal attenuating noticeably with increasing distance.

The results, reported in Fig.~\ref{fig:barplot_2}, show a clear improvement in both \gls{dl} and \gls{ul} spectral efficiency as the \gls{wab} node is moved closer to the \gls{ue}, highlighting the influence of access link quality.  
Despite the limited transmit power of the WAB-gNB, the end-to-end \gls{ul} spectral efficiency of the \gls{wab} system surpasses that achieved by the FR2-only \gls{cpe}. These findings confirm that the \gls{wab} architecture can effectively mitigate the typical limitations affecting FR2 \gls{ul} transmissions~\cite{hfcl}, delivering enhanced performance even in challenging \gls{nlos} indoor scenarios.

In the \gls{dl} direction, the end-to-end spectral efficiency of the \gls{wab} setup remains lower than that of the FR2-only \gls{cpe}. This limitation is primarily due to the prototype nature of the WAB-gNB used in our testbed.  
However, we expect that replacing the current FR1 configuration with a more powerful, commercial-grade FR1 system would significantly enhance downlink performance, potentially allowing the \gls{wab} architecture to outperform the FR2-only configuration even in the \gls{dl} direction.

\section{Conclusions \& Future Work}
\label{sec:conclusion}
This paper introduced \gls{wab} as a promising enabler for mobile base stations and flexible wireless backhaul, as well as an effective approach for exploiting FR2 communications. We presented a \gls{wab} testbed that addresses the current lack of experimental validation in the literature and provides a practical foundation for further research.

Through an extensive experimental campaign conducted in both mobile and \gls{o2i} scenarios, the proposed framework, based on commercial hardware and open-source software, demonstrated the feasibility and benefits of multi-band \gls{wab} operation. In particular, the combination of FR1 access and FR2 backhaul proved effective in extending coverage and improving reliability under mobility and challenging propagation conditions. The use of standard 5G interfaces, together with the flexibility and multi-technology backhaul support of the \gls{wab} architecture, makes it a compelling solution for vehicular and next-generation wireless networks.

We are currently working on extending the proposed testbed to better assess backhaul handover, improve interference and resource management strategies, and integrate additional backhaul technologies to further enhance scalability and performance.


\vspace{-0.2cm}
\bibliographystyle{IEEEtran}
\bibliography{bibliography.bib}

\end{document}